\input epsf.tex
\def\DESepsf(#1 width #2){\epsfxsize=#2 \epsfbox{#1}}

\documentclass{ws-p8-50x6-00}

\begin{document}

\title{NEUTRALINO PROTON CROSS SECTION AND DARK MATTER DETECTION}

\author{R. Arnowitt, B. Dutta and Y. Santoso}

\address{Center For Theoretical Physics, Department of Physics, Texas A$\&$M University,
College Station TX 77843-4242, USA} 


\maketitle

\abstracts{We consider the neutralino proton cross section for detection of Milky Way
dark matter for a number of supergravity models with gauge unification at
the GUT scale: models with universal soft breaking (mSUGRA), models with
nonuniversal soft breaking, and string inspired D-brane models. The
parameter space examined includes $m_{1/2}<$1 TeV and $\tan \beta <$ 50, and the
recent Higgs bound of $m_h >$114 GeV is imposed. (For grand unified models, this
bound is to be imposed for all $\tan \beta$.) All coannihilation effects
are included as well as the recent NLO corrections to $b\rightarrow s\gamma$ for large
$\tan \beta$, and coannihilation effects are shown to be sensitive to $A_0$ for
large $\tan \beta$. In all models, current detectors are sampling parts of the
paramater space i. e. $\tan \beta \stackrel{>}{\sim}$ 25 for mSUGRA, 
$\tan \beta \stackrel{>}{\sim}$ 7 for
nonuniversal models, and $\tan \beta \stackrel{>}{\sim}$ 20 for D-brane models. Future detectors
should be able to cover almost the full parameter space for $\mu >$ 0. For $\mu<$ 0, 
cancellations can occur for $m_{1/2} \stackrel{>}{\sim}$ 450 GeV, allowing the cross
sections to become $\stackrel{<}{\sim} 10^{-10}$ pb for limited ranges of $\tan
\beta$. (The positions of these cancellations are seen to be sensitive to the
value of $\sigma_{\pi N}$.) In this
case, the gluino and squarks lie above 1 TeV, but still should be
accessible to the LHC if $m_{1/2} <1$ TeV.}

\section{Introduction} The existance of dark matter, which makes up about 30$\%$ of all the
matter and energy in the universe, is well documented astronomically. However, what it is made
of is unknown, and there have been many theoretical suggestions: wimps, axions, machos, etc.
The Milky way consists of perhaps 90$\%$ dark matter, and so is a convenient
``laboratory'' for
the study of dark matter, particularly by direct detection by terrestial detectors. We
consider here the case of supersymmetric (SUSY) wimp dark matter and its detection by
scattering by nuclear targets. In SUSY models with R-parity invariance, the wimp is almost
always the lightest neutralino, $\tilde\chi^0_1$, and for heavy nuclei, the spin independent
scattering dominates the  cross section. Since then neutron and proton cross sections in the
nuclei are nearly equal, it is possible to extract the $\tilde\chi^0_1-p$ cross section, 
$\sigma_{\tilde\chi^0_1-p}$, from any data (subject, of course, to astronomical
uncertainties). Current detectors (DAMA, CDMS, UKDMC) are sensitive to cross sections
\begin{equation}
\sigma_{\tilde\chi^0_1-p}\stackrel{>}{\sim} 1\times 10^{-6} \, {\rm pb}
\end{equation} with perhaps an improvement on this by one or two orders of magnitude in the
near future. More long range, future detectors (GENIUS, Cryoarray) plan on a significant
increase in sensitivity, i.e. down to
\begin{equation}
\sigma_{\tilde\chi^0_1-p}\stackrel{>}{\sim} (10^{-9}-10^{-10}) \, {\rm pb}
\end{equation} We discuss here how such sensitivities might relate to what is expected from
supersymmetry models.

We consider here three SUSY models based on grand unification of the gauge coupling constants
at the GUT scale of $M_G\cong 2\times 10^{16}$ GeV:
\begin{enumerate}
\item
Minimal Supergravity GUT Models (mSUGRA)\cite{SUGRA}. Here there are universal soft breaking
masses occurring at scale $M_G$.
\item
Non-universal Soft Breaking Models\cite{berezinsky}. Here the first two generation of squarks
and sleptons soft breaking masses are kept universal (to suppress flavor changing neutral
currents) and the gaugino masses are universal at $M_G$, while nonuniversalities are allowed in
the Higgs soft breaking masses and the third generation squark and sleptons masses at $M_G$.
\item
D-brane String Models (based on type IIB Orientifolds)\cite{munoz,brhlik}. Here the $SU(2)_L
$doublet scalar masses are different from the singlet masses at $M_G$, and the gaugino masses
are similarly not degenerate.
\end{enumerate}
The three types of models have varying amount of complexity in the soft breaking parameters,
and while the first two models arise from natural phenomenological considerations in
supergravity theory, there are also string models that can realise such soft breaking
patterns. Though physically very different, all the models turn out to lead to qualitatively
similar results: Current detectors are sensitive to a significant part of the SUSY parameter
space, and future detectors should be able to cover most of the parameter space except for some
special regions where accidental cancellations can occur which make
$\sigma_{\tilde\chi^0_1-p}$ anomalously small. Thus dark matter experiments offer significant
tests of supersymmetry over the same time scale (the next 5-10 years) that accelerator
experiments will.

While each of the above models contain a number of unknown parameters, theories of this type
can still make relevant predictions for two reasons: (i) they allow for radiative breaking of
$SU(2)\times U(1)$ at the electroweak scale (giving a natural explanation of the Higgs
mechanism), and (ii) along with calculating
$\sigma_{\tilde\chi^0_1-p}$, the theory can calculate the relic density of
$\tilde\chi^0_1$, i.e
$\Omega_{\tilde\chi^0_1}=\rho_{\tilde\chi^0_1}/\rho_c$ where
$\rho_{\tilde\chi^0_1}$ is the relic mass density of  $\tilde\chi^0_1$ and
$\rho_c=3 H_0^2/8\pi G_N$ ($H_0$ is the Hubble constant and $G_N$ is the Newton constant).
Both of these greatly restrict the parameter space. In general one has
$\Omega_{\tilde\chi^0_1}h^2\sim(\int^{x_f}_0 dx\langle\sigma_{\rm ann}v\rangle)^{-1}$ (where
$\sigma_{\rm ann}$ is the neutralino annihilation cross section in the early universe, $v$ is
the relative velocity,
$x_f=kT_f/m_{\tilde\chi^0_1}$,
$T_f$ is the freeze out temperature, $\langle...\rangle$ means thermal average and $h=H_0/100$
km s$^{-1}$Mpc$^{-1}$). The fact that these conditions can naturally be satisfied for reasonable parts
of the SUSY parameter space represents a significant success of the SUGRA models.

In the following we will assume $H_0=(70\pm 10)$km s$^{-1}$Mpc$^{-1}$ and matter (m) and
baryonic (b) relic densities of $\Omega_m=0.3\pm 0.1$ and
$\Omega_b=0.05$. Thus $\Omega_{\tilde\chi^0_1}h^2=0.12\pm 0.05$. The calculations given below
allow for a 2$\sigma$ spread, i.e. we take \cite{com} 
\begin{equation} 0.02\leq \Omega_{\tilde\chi^0_1}h^2\leq 0.25. 
\end{equation}

It is clear that accurate determinations of the dark matter relic density will greatly
strengthen the theoretical predictions, and already, analyses using combined data from the
CMB, large scale structure, and supernovae data suggests that the correct value of the relic
density lies in a relatively narrow band in the center of the region of Eq.
(3)\cite{fukugita}. We will
here, however, use the conservative range given in Eq. (3).

\section{Theoretical Analysis} In order to get accurate predictions of the maximum and minimum
cross sections for a given model, it is necessary to include a number of theoretical
corrections. We list here the main ones: (i) In relating the theory at $M_G$ to phenomena at
the electroweak scale, the two loop gauge and one loop Yukawa renormalization group equations
(RGE) are used, iterating to get a consistent SUSY spectrum. (ii) QCD RGE corrections are
further included below the SUSY breaking scale for contributions involving light quarks. (iii)
A careful analysis of the light Higgs mass $m_h$ is necessary (including two loop and pole
mass corrections) as the current LEP limits impact sensitively on the relic
density analysis. (iv) L-R mixing terms are included in the sfermion (mass)$^2$ matrices
since they produce important effects for large $\tan \beta$ in the third generation. (v) One
loop corrections are included to
$m_b$ and
$m_{\tau}$ which are again important for large $\tan \beta$. (vi) The experimental bounds on the
$b\rightarrow s\gamma$ decay put significant constraints on the SUSY parameter space and
theoretical calculations here include the leading order (LO) and NLO corrections. We have not
in the following imposed
$b-\tau$ (or $t-b-\tau$) Yukawa unification or proton decay constraints as these depend
sensitively on unknown post-GUT physics. For example, such constraints do not naturally occur
in the string models where
$SU(5)$ (or $SO(10)$) gauge symmetry is broken by Wilson lines at $M_G$ (even though grand
unification of the gauge coupling constants at $M_G$ for such string models is still required).

All of the above corrections are now under theoretical control. In particular, the
$b\rightarrow s\gamma$ SUSY NLO corrections for large $\tan \beta$ have recently been calculated
\cite{gam,car}. We find that the NLO corrections give
significant contributions for large $\tan \beta$ for $\mu >$0. (We use here Isajet sign conventions
for the $\mu$ parameter.) There have been a number of calculations of
$\sigma_{\tilde\chi^0_1-p}$ given in the literature 
\cite{bottino,bottino1,falk,ellis,falk1,aads,aad}, and we find we are in general numerical
agreement in those regions of parameter space where the authors have taken into account the
above corrections.

Accelerator bounds significantly limit the SUSY parameter space. As pointed out in
\cite{nano}, the LEP bounds on the Higgs mass has begun to make a significant impact on dark
matter analyses. Since at this time it is unclear whether the recently observed LEP events
\cite{lephiggs} represent a Higgs discovery, we will use here the current LEP lower bound of 114
GeV
\cite{igo}. There are still some remaining errors in the theoretical calculation of the
Higgs mass, however, as well as uncertainty in the $t$-quark mass, and so we will
conservatively assume here for the light Higgs ($h$) that $m_h > 110$ GeV for
all $\tan \beta$. (For the MSSM, the Higgs mass constraint is significant only
for $\tan \beta \stackrel{<}{\sim} 9$ (see e.g. Igo-Kemenes\cite{igo}) as $Ah$ production with
$m_A \cong m_Z$ can be confused with $Zh$ production. However, in GUT models
radiative breaking eliminates such regions of parameter space and the LEP
constraint operates for all $\tan \beta$.) LEP data also produces a bound on the lightest chargino
($\tilde\chi^{\pm}_1$) of 
$m_{\tilde\chi^{\pm}_1}> 102$ GeV \cite{lepchar}. For $b\rightarrow s\gamma$ we assume an
allowed range of $2\sigma$ from the CLEO data \cite{cleo}:
\begin{equation} 1.8\times 10^{-4}\leq B(B\rightarrow X_s\gamma)\leq 4.5 \times 10^{-4}
\end{equation} The Tevatron gives a bound of $m_{\tilde g}\geq 270$ GeV( for
$m_{\tilde q}\cong m_{\tilde g}$)\cite{comm1}.

Theory allows one to calculate the $\tilde\chi^0_1$-quark cross section and we follow the
analysis of \cite{ellis2} to convert this to $\tilde\chi^0_1-p$ scattering. For this one needs
the $\pi-N$ sigma term,
\begin{equation}
\sigma _{\pi N}={1\over 2}(m_u+m_d)\langle p|{\bar u} u+{\bar d}d|p\rangle,
\end{equation}
$\sigma_0=\sigma _{\pi N}-(m_u+m_d)\langle p|{\bar s} s|p\rangle$ and the quark mass ratio
$r=m_s/{(1/2)(m_u+m_d)}$. We use here $\sigma_0= 30$ MeV
\cite{bottino1}, and $r=24.4\pm 1.5$\cite{leutwyler}. Recent analyses, based on new $\pi-N$
scattering data gives $\sigma _{\pi N}\cong 65$ MeV\cite{mgo,pas}. Older 
$\pi-N$ data gave $\sigma _{\pi N}\cong 45$ MeV\cite{gs}.  We will use in most of the analysis
below the larger number. If the smaller number is used, it would have the overall effect in
most of the parameter space of reducing  $\tilde\chi^0_1-p$ by about a factor of 3. However, in
the special situation for $\mu < $0, where there is a cancellation of matrix elements, the
choice of $\sigma _{\pi N}$ produces a more subtle effect, and we will exhibit there results
from both values.

\section{mSUGRA model} We consider first the mSUGRA model where the most complete analysis has
been done. mSUGRA depends on four parameters and one sign:
$m_0$ (universal scalar mass at $M_G$), $m_{1/2}$ (universal gaugino mass at
$M_G$), $A_0$ (universal cubic soft breaking mass), $\tan \beta=\langle H_2\rangle/{\langle
H_1\rangle}$ (where $\langle H_{(2,1)}\rangle$ gives rise to (up, down) quark masses) and
$\mu/|\mu|$ (where $\mu$ is the Higgs mixing parameter in the superpotential,
$W_{\mu}=\mu H_1H_2$). One conventionally restricts the range of these parameters by
``naturalness" conditions and in the following we assume $m_0\leq 1$ TeV, $m_{1/2}\leq 1$ TeV
(corresponding to $m_{\tilde g}\leq 2.5$ TeV,
$m_{\tilde \chi^0_1}\leq 400$ GeV), $|A_0/m_{1/2}|\leq 4$, and 2$\leq \tan \beta\leq$ 50. Large
$\tan \beta$ is of interest since SO(10) models imply $\tan \beta\geq 40$ and also
$\sigma_{\tilde\chi^0_1-p}$ increases with $\tan \beta$.
$\sigma_{\tilde\chi^0_1-p}$ decreases with $m_{1/2}$ for large $m_{1/2}$.

The maximum value of $\sigma_{\tilde\chi^0_1-p}$ arises then for large $\tan \beta$ and small
$m_{1/2}$. This can be seen in Fig.1 where ($\sigma_{\tilde\chi^0_1-p}$)$_{\rm max}$ is
plotted vs. $m_{\tilde \chi^0_1}$ for $\tan \beta$=20, 30, 40 and  50. Fig. 2 shows
$\Omega_{\tilde\chi^0_1}h^2$  for $\tan \beta=30$ when the cross section takes on  its maximum
value. Current detectors obeying Eq (1) are then sampling the parameter space
for large $\tan \beta$, small $m_{\tilde \chi^0_1}$ and small $\Omega_{\tilde\chi^0_1}h^2$ i.e 
\begin{equation} \tan\beta\stackrel{>}{\sim}25,\,m_{\tilde
\chi^0_1}\stackrel{<}{\sim} 90 {\rm GeV},\,\Omega_{\tilde\chi^0_1}h^2\stackrel{<}{\sim} 0.1
\end{equation} 
Further, as can be seen from Fig. 3, $m_h$ does indeed exceed the current LEP bound over
this entire region.

\begin{figure}[htb]
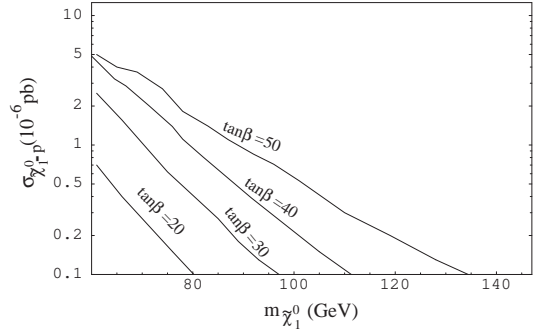

\centerline{ \DESepsf(aads20304050.epsf width 7 cm) }
\bigskip
\bigskip
\caption{\label{fig1}  ($\sigma_{\tilde\chi^0_1-p}$)$_{\rm max}$ for mSUGRA obtained by
varying $A_0$ and $m_0$ over the parameter space for  $\tan \beta =20$, 30, 40,
and 50$^{14}$. The
relic density constraint, Eq.(3) has been imposed.}
\end{figure}

\begin{figure}[htb]
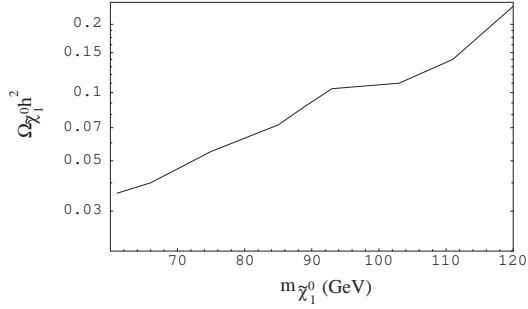

\centerline{ \DESepsf(aads30om.epsf width 7 cm) }
\bigskip
\bigskip
\caption {\label{fig2} $\Omega_{\tilde{\chi}_{1}^{0}}h^{2}$ for mSUGRA when
$(\sigma_{\tilde{\chi}_{1}^{0}-p})$ takes on its  maximum value for $\tan \beta
=30$$^{14}$.}
\end{figure}    
\begin{figure}[htb]
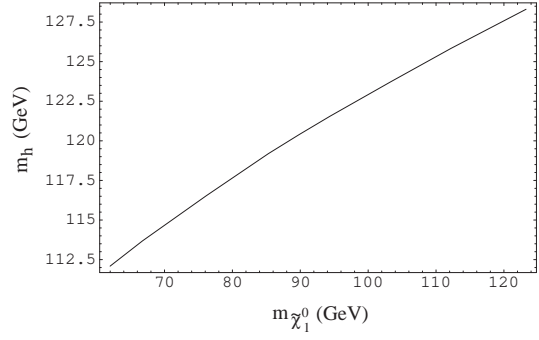

\centerline{ \DESepsf(aads30higgs.epsf width 7 cm) }
\caption {\label{fig4} m$_h$ for mSUGRA as a function of $m_{\tilde\chi^0_1}$ for
$\tan \beta=$30,  when $\sigma_{\tilde\chi^0_1-p}$ takes on it maximum
value$^{14}$.}
\end{figure}

As discussed in \cite{falk}, coannihilation effects in the early universe can significantly
influence the relic density calculation. To discuss the minimum cross section,
it is convenient then to
consider first 
$m_{\tilde \chi^0_1}\stackrel{<}{\sim} 150$ GeV ($m_{1/2}\leq 350$) where no coannihilation
occurs. The minimum cross section occurs for small $\tan \beta$. From Fig.4 one sees
\begin{equation}
\sigma_{\tilde\chi^0_1-p}\stackrel{>}{\sim} 1\times 10^{-9} {\rm pb};\,
 m_{\tilde \chi^0_1}\stackrel{<}{\sim} 140 {\rm GeV};\,\tan \beta=6
\end{equation} which would be accessible to detectors that are currently being planned (e.g.
GENIUS).

\begin{figure}[htb]
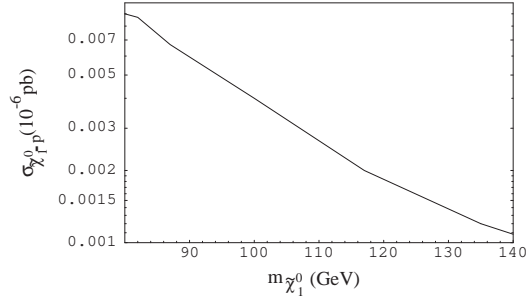

\centerline{ \DESepsf(aadtan6uninew.epsf width 7 cm) }
\caption {\label{fig41} ($\sigma_{\tilde\chi^0_1-p}$)$_{\rm min}$ for mSUGRA 
is plotted as a function of
$m_{\tilde\chi^0_1}$ for $\mu>$0, $\tan \beta=6$.}
\end{figure}

For larger $m_{\tilde \chi^0_1}$, i.e. $m_{1/2}\stackrel{>}{\sim} 350$ the phenomena of
coannihilation can occur in the relic density analysis since the light stau, $\tilde \tau_1$,
(and also $\tilde e_R$, $\tilde \mu_R$) can become degenerate with the $\tilde\chi^0_1$. The
relic density constraint can then be satisfied in narrow corridor of $m_0$ of width 
$\Delta m_0\stackrel{<}{\sim}25$ GeV, the value of $m_0$ increasing as $m_{1/2}$
increases and this was examined for low and intermediate $\tan \beta$ in
\cite{falk}. Since $m_0$ and $m_{1/2}$ increase as one progresses up the corridor, 
$\sigma_{\tilde\chi^0_1-p}$ will generally decrease.

We consider first the case of $\mu >$ 0 \cite{ads}. Coannihilation effects generally begin for
$m_{1/2}\stackrel{>}{\sim}400$ GeV ($m_{\tilde \chi^0_1}\stackrel{>}{\sim}160$ GeV), and it is
of interest to see what occurs for large $\tan \beta$. For large $\tan \beta$, there is only a
coannihilation region left in the parameter space, and the allowed regions, exhibiting the
allowed narrow corridors of parameter space are shown in Fig. 5 for $\tan \beta=40$. In this
domain the lightest stau ($\tilde \tau_1$) is the lightest slepton due to the large L-R mixing
in the (mass$^2$) matrix, and so dominates the conanihilation effects. We note that the
allowed corridors are sensitive to $A_0$, and large $A_0$ can allow large $m_0$ as $m_{1/2}$
increases. The thickness of the allowed corridors also decrease as
$A_0$ increases. There is also a lower bound on $m_{1/2}$ for the allowed regions due to the
$b\rightarrow s\gamma$ constraint, this bound decreasing with increasing
$A_0$. (We note that this lower bound is sensitive to the NLO corrections discussed in Sec. 2
above.)
\begin{figure}[htb]
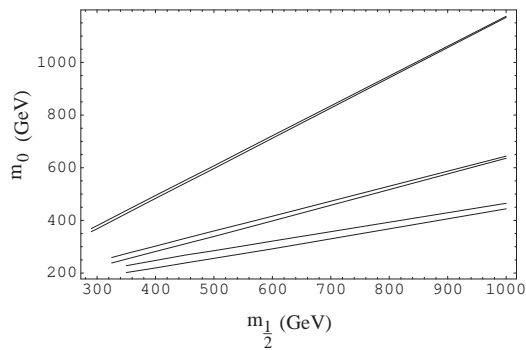

\centerline{ \DESepsf(aadcoan40new.epsf width 7 cm) }
\caption {\label{fig9} Allowed corridors for mSUGRA in the $m_0-m_{1/2}$ plane
satisfying
the relic density constraint of Eq(3) for $\mu>$0, $\tan \beta=40$ (from bottom to top)
$A_0=$ $m_{1/2}$, 2$m_{1/2}$, 4$m_{1/2}$ $^{27}$.
 }
\end{figure}  
Since larger $A_0$ allows for larger $m_0$ in the coannihilation region, the
scattering cross section is a decreasing function of $A_0$. This is shown in
Fig. 6 where
$\sigma_{\tilde\chi^0_1-p}$ is plotted as a function of $m_{1/2}$ for $\tan \beta=40$ and
$A_0 =$ $2m_{1/2}$, $4m_{1/2}$.

\begin{figure}[htb]
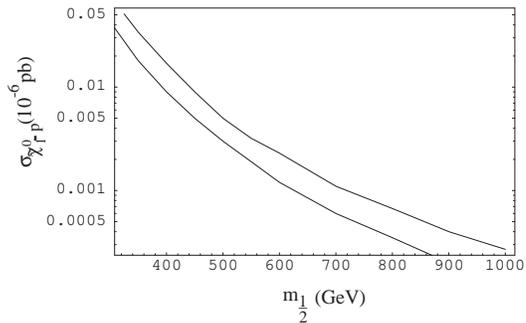

\centerline{ \DESepsf(aadcoan4024new.epsf width 7 cm) }
\caption {\label{fig91}$\sigma_{\tilde\chi^0_1-p}$  as a function of
$m_{1/2}$ for mSUGRA, $\mu>$0, $\tan \beta=40$ and
$A_0 =$ $2m_{1/2}$ (upper curve), $4m_{1/2}$ (lower curve).
 }
\end{figure}

We consider next $\mu <$ 0. As discussed in \cite{ellis}, for low and intermediate $\tan
\beta$, an
accidental cancellation can occur in the heavy and light Higgs amplitudes in the
coannihilation region which can greatly reduce 
$\sigma_{\tilde\chi^0_1-p}$. We investigate here what happens at larger $\tan
\beta$, and what is
the domain over which this cancellation occurs. In Fig. 7  we have plotted
$\sigma_{\tilde\chi^0_1-p}$ in the large $m_{1/2}$ region, for $\tan \beta = 6$ (short dash),
10(solid), 20(dot-dash), and 25(dashed). One sees that the cross section dips sharply for
$\tan \beta = 10$, reaching a minimum at $m_{1/2}\stackrel{\sim}{=}$725 GeV, and then rises.
Similarly, for $\tan \beta = 20$, the minimum occurs at 
$m_{1/2}\stackrel{\sim}{=}$ 830 GeV while for $\tan \beta = 25$ at $m_{1/2}\stackrel{\sim}{=}$ 950
GeV. As a consequence, 
$\sigma_{\tilde\chi^0_1-p}$ will fall below the sensitivity of planned future detectors for
$m_{1/2}>$450 GeV in a restricted region of $\tan \beta$, i.e.
\begin{equation}
  \sigma_{\tilde\chi^0_1-p} < 1 \times 10^{-10}\,\,{\rm for}\, 450\,{\rm GeV}
   < m_{1/2} < 1\,{\rm TeV};\, 5 \stackrel{<}{\sim}\tan\beta\stackrel{<}{\sim}30;\, 
   \mu<0.
\end{equation} 

At the minima, the cross sections can become quite small, e.g. $1
\times10^{-13}$pb, without major fine tuning of
parameters, corresponding to almost total cancellation. Further,
the widths of the minima at fixed $\tan \beta$ are fairly broad. While in this domain proposed
detectors would not be able to observe Milky Way wimps, mSUGRA would imply that the squarks
and gluinos then would lie above 1 TeV, but at masses that would still be accesible to the LHC.
Also mSUGRA implies that this phenomena can occur only in a restricted range of
$\tan \beta$,
and for $\mu <$0, so there would still be a number of cross checks of the theory.
\begin{figure}[htb]
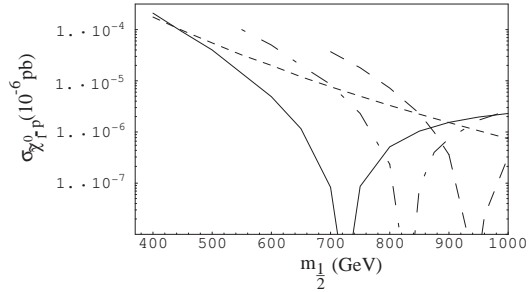

\centerline{ \DESepsf(aadcoan61020.epsf width 7 cm) }
\caption {\label{fig5} $(\sigma_{\tilde{\chi}_{1}^{0}-p})$ for mSUGRA  for $\mu<0$, $A_0=$1500
GeV for $\tan \beta = 6$ (short dash), 10 (solid), 20 (dot-dash), 25 (dashed).  Note that the
$\tan \beta = 6$ curve terminates at low $m_{1/2}$  due to the Higgs mass constraint, and the
other curves terminate at low $m_{1/2}$ due to the $b\rightarrow s\gamma$ constraint.}
\end{figure}

In Sec. 2 it was pointed out that there was considerable uncertainty in the properties of the
nucleon, particularly in the value of the $\sigma_{\pi N}$. In the above curves we have used
$\sigma_{\pi N}=$65 MeV. Fig. 8 gives a comparison
 for $\tan \beta = 10$ of this choice (solid curve) with the parameters of
 \cite{ellis} (dashed curve)
where $\sigma_{\pi N}=$45 MeV is used.  One sees that the position of the minimum at
$m_{1/2}=$ 725 GeV is shifted to 600 GeV, with similar shifts occuring for the other values of
$\tan \beta$. Thus the cancellations ocurring in this region are quite sensitive to the properties
of the proton.
\begin{figure}[htb]
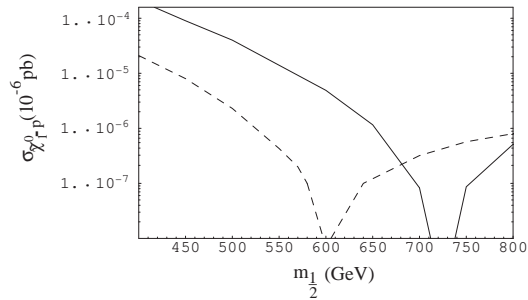

\centerline{ \DESepsf(aadcoanellis10.epsf width 7 cm) }
\caption {\label{fig8} $(\sigma_{\tilde{\chi}_{1}^{0}-p})$ for mSUGRA for $\mu <$ 0 
$\tan \beta = 10$, $\sigma_{\pi N}=$65 GeV (solid), and the parameters of
$^{12}$ (dashed) where
$\sigma_{\pi N}=$45 GeV was used.}
\end{figure}
\section{Nonuniversal SUGRA Models}

In most discussions of SUGRA models with nonuniversal soft breaking terms, the universality of
the the soft breaking masses at $M_G$ of the first two generations of squarks and sleptons is
maintained to suppress flavor changing neutral currents. However, one may allow both the Higgs
masses and the third generation squark and slepton masses to become nonuniversal at
$M_G$. One can parameterize this situation at $M_G$ as follows:

\begin{eqnarray} m_{H_{1}}^{\ 2}&=&m_{0}^{2}(1+\delta_{1}); 
\quad m_{H_{2}}^{\ 2}=m_{0}^{2}(1+ \delta_{2});\\\nonumber m_{q_{L}}^{\
2}&=&m_{0}^{2}(1+\delta_{3}); \quad m_{t_{R}}^{\ 2}=m_{0}^{2}(1+\delta_{4});
\quad m_{\tau_{R}}^{\ 2}=m_{0}^{2}(1+\delta_{5}); 	\\\nonumber m_{b_{R}}^{\
2}&=&m_{0}^{2}(1+\delta_{6}); \quad m_{l_{L}}^{\ 2}=m_{0}^{2}(1+\delta_{7}).
\label{eq18}
\end{eqnarray} where $q_{L}\equiv (\tilde t_L, \tilde b_L)$ squarks, $l_{L}\equiv (\tilde
\nu_\tau, \tilde \tau_L)$ sleptons,
 etc. and $m_0$ is the universal mass for the first two generations of squarks and sleptons.
The $\delta_i$ are the deviations from universality (and if one were to impose SU(5) or SO(10)
symmetry one would have $\delta_3$=$\delta_4$=
$\delta_5$, and
$\delta_6$=$\delta_7$.) In the following we limit the $\delta_i$ to obey:
\begin{equation}
                 -1 \leq \delta_i \leq +1
\end{equation} and maintain gauge coupling constant unification and gaugino mass unification
at $M_G$.

While there are a large numbers of new parameters, one can get an understanding of what effect
they produce from the following. The neutralino ${\tilde\chi^0_1}$ is a mixture of gaugino
(mostly Bino) and higgsino parts:
\begin{equation}
      {\tilde\chi^0_1}= \alpha \tilde W_3 + \beta \tilde B + \gamma \tilde H_1 +
       \delta \tilde H_2
\end{equation} 
The dominant spin independent $\sigma_{\tilde\chi^0_1-p}$ cross section is
proportional to the interference between the gaugino and higgino amplitudes, and this
interference is largely governed by the size of $\mu^2$. As $\mu^2$ decreases, the
interference increases, and hence $\sigma_{\tilde\chi^0_1-p}$ increases. Radiative breaking of
$SU(2)\times U(1)$ determines the value of $\mu^2$ at the electroweak scale. To see the general
nature of the effects of nonuniverality, we consider low and intermediate $\tan \beta$ where an
analytic form exists for
$\mu^2$ (see e.g. Arnowitt and Nath \cite{berezinsky}):
\begin{eqnarray}
\mu^2&=&{t^2\over{t^2-1}}\left[({{1-3 D_0}\over 2}-{1\over
t^2})+{{1-D_0}\over2}(\delta_3+\delta_4)\right.\\\nonumber 
&-&\left.{{1+D_0}\over2}(\delta_2+{\delta_1\over
t^2})\right]m_0^2+{\rm {universal\,parts\,+\,loop \, corrections}}.
\end{eqnarray} 
Here $t=\tan\beta$ and $D_0\cong1-(m_t/200{\rm GeV} \sin\beta)^2 \leq 0.2$  (Note
that the Higgs and squark nonuniversalties enter coherrently, roughly in the combination
$\delta_3+\delta_4-\delta_2$.) We see from Eq.(12) that $\mu^2$ is reduced, and hence
$\sigma_{\tilde\chi^0_1-p}$ increased for 
$\delta_3$, $\delta_4$, $\delta_1<0$,
$\delta_2>0$, and $\mu^2$ is increased for $\delta_3$, $\delta_4$, 
$\delta_1>0$, $\delta_2<0$. Thus one can get significantly larger cross sections in the
nonuniversal models with the first choice of signs for the $\delta_i$, and one can reduce the
cross sections (though not by such a large amount) with the second choice.  In general this
would allow one to significantly lower the value of $\tan \beta$ from the rquirement 
of $\tan \beta\stackrel{>}{\sim} 25$ for mSUGRA to come within the range of
current detectors. The matter
is complicated, however, by the experimental 
Higgs mass constraint (that $m_h > 110$) since theoretically the Higgs mass is
small for low $\tan \beta$. We find, however, that one can reduce $\tan \beta$ to 7, and still maintain $m_h >
110$ GeV. This is exhibited in Fig. 9, where the maximum cross section is plotted as a
function of 
$m_{\tilde\chi^0_1}$. The nonuniversal cross sections can be a factor of 10 or more greater than
the corresponding universal ones allowing for these much lower values of $\tan
\beta$. 

\begin{figure}[htb]
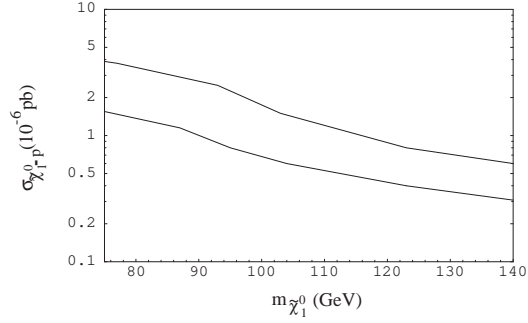

\centerline{ \DESepsf(aadtan712new.epsf width 7 cm) }
\caption {\label{fig81} Maximum $\sigma_{\tilde\chi^0_1-p}$ for nonuniversal model for $\mu>$0 with 
$\delta_3$, $\delta_4$,
$\delta_1 <$ 0, $\delta_2 > $0. Lower curve is for $\tan \beta = 7$ and the upper
curve is for $\tan \beta = 12$.}
\end{figure}
As in mSUGRA, the minimum cross sections occur for the largest $m_{1/2}$ and smallest
$\tan \beta$, and so they occur in the coannihilation region. We consider here the case
where only the Higgs masses are nonuniversal i.e.
$\delta_{1,2} \neq0$ (the other $\delta_i$ set to zero). Results then are similar to the
mSUGRA case. For $\mu>0$ we find
\begin{equation}
\sigma_{\tilde\chi^0_1-p}\stackrel{>}{\sim} 0.3(1)\times 10^{-10} {\rm pb};\,\,\,{\rm
for}\,\,\,
 m_{1/2} \leq 1(0.6)\, {\rm TeV}
\end{equation}  
For $\mu<$0, one can again get a cancelation reducing the cross sections below
$1\times10^{-10}$ pb for $m_{1/2} > 450$ GeV.  As in mSUGRA these can produce sharp minima in
the cross sections in the region $\tan \beta = $10-25.
\section{D-Brane Models} 

Recent advances in string theory has stimulated again the building of string inspired models.
We consider here models based on Type $IIB$ orientifolds where the full $D=10$ space was
compactified on a six torus $T^6$ \cite{munoz}. These models can contain 9-branes and 5-branes which
can be used to embed the Standard Model. We consider here a model in which 
$SU(3)_C \times U(1)_Y$ is associated with one set of 5-branes, $5_1$ and 
$SU(2)_L$ is associated with a second intersecting set $5_2$ \cite{brhlik}. Strings beginning and ending
on $5_1$ will have massless modes carrying the $SU(3)_C \times U(1)_Y$ quantum numbers (i.e.
the R quarks and R leptons), while strings starting on $5_2$ and ending on $5_1$ will have
massless modes carrying the joint quantum numbers of the two branes (i.e. the quark, lepton
and Higgs doublets). This then leads to the following soft breaking pattern at $M_G$:
\begin{eqnarray}
\tilde m_1&=&\tilde m_3=-A_0=\sqrt{3} \cos\theta_b \Theta_1
e^{-i\alpha_1}m_{3/2}\\\nonumber
\tilde m_2&=&\sqrt{3} \cos\theta_b (1-\Theta_1^2)^{1/2}m_{3/2}
\end{eqnarray} 
where and $\tilde m_i$ are the gaugino masses, and 
\begin{eqnarray} 
m_{12}^2&=&(1-3/2 \sin^2\theta_b)m_{3/2}^2\, \, {\rm for} \,\,
q_L,\,l_L,\,H_1,\,H_2\\\nonumber 
m_{1}^2&=&(1-3 \sin^2\theta_b)m_{3/2}^2\,\, {\rm for}\,\,
u_R,\,d_R,\, e_R.
\end{eqnarray} 
Thus the $SU(2)$ doublets are all degenerate at $M_G$ but are different from the
singlets. We note Eq. (15) implies $\theta_b<$0.615

This model was initially studied to examine its CP violating properties. However, it was
subsequently seen that the experimental constraint of the electron electric dipole moment led
to a serious fine tuning problem at $M_G$ unless $\tan
\beta\stackrel{<}{\sim}3-5$ \cite{aadcp}. Since we
are interested here in larger $\tan\beta$, we will set the CP violating phases to zero.

In general, Eq (15) shows that $\sigma_{\tilde{\chi}_1^0-p}$ is an increasing function of
$\theta_b$ since the squark and slepton masses decrease with increasing
$\theta_b$. Thus the maximum cross sections will arise from large $\theta_b$ and large
$\tan \beta$. This is illustrated in Fig. 10, where $\sigma_{\tilde\chi^0_1-p}$
 is plotted as a function of $m_{\tilde \chi^0_1}$ for $\mu >0$ for $\tan \beta = $20,
and $\theta_b = 0.2$. Thus we see that current detectors obeying the bound of Eq. (1) are
sampling the parameter space for
\begin{equation}
                     \tan\beta\stackrel{>}{\sim}20
\end{equation} 
We note that when $\tan \beta$ is close to its minimum value, $m_{\tilde \chi^0_1}$
is also close to it's current LEP bound of $m_{\tilde \chi^0_1} > $37 GeV
\cite{neutmass,commx}.
\begin{figure}[htb]
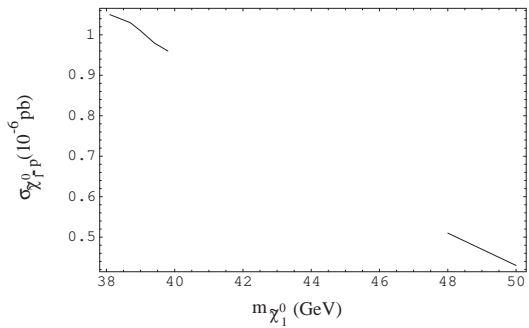

\centerline{ \DESepsf(aadtan20dark.epsf width 7 cm) }
\caption {\label{fig811} $\sigma_{\tilde\chi^0_1-p}$ for D-brane model for $\mu > 0$, 
$\theta_b=0.2$ and $\tan\beta = 20$. The gap in the curve is due to excessive early universe
annihilation through s-channel $Z$ and $h$ poles.}
\end{figure}
The minimum value of $\sigma_{\tilde\chi^0_1-p}$ will occur at low $\theta_b$,
low $\tan\beta$,
and large $m_{3/2}$ (i.e. large $m_{\tilde \chi^0_1}$). In the large 
$m_{\tilde \chi^0_1}$ region, coannihilation can occur between the sleptons and the neutralino
in a fashion similar to the SUGRA models, with the effective slepton $m_0$ parameter and
effective neutralino $m_{1/2}$ parameter being given by
\begin{eqnarray} 
m_0^2 &=&(1 - 3 \sin^2\,\theta_b)\,m_{3/2}^2\\ \nonumber 
m_{1/2} &=& \sqrt{3}\, \cos\theta_b\, \Theta_1\, m_{3/2}
\end{eqnarray} 

\begin{figure}[htb]
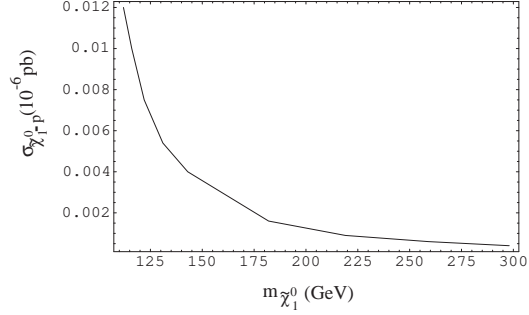

\centerline{ \DESepsf(aadcoandbranenew.epsf width 7 cm) }
\caption {\label{fig141} Minimum $\sigma_{\tilde\chi^0_1-p}$ for the D-brane model for $\mu < 0$ and
$\tan\beta = 6$.}
\end{figure}
Fig. 11 exhibits the minimum cross section for $\mu > 0$ as a function of the
neutralino mass. One sees that
\begin{equation}
 \sigma_{\tilde\chi^0_1-p}\stackrel{>}{\sim} 1 \times10^{-9}\,\,{\rm pb}
 \,\,{\rm  for}\,\, \mu > 0
\end{equation} 
which is accessible to planned detectors. We note also that coannihilation is
possible between the light chargino and neutralino. However, this occurs for only a very small
region of parameter space.

As in mSUGRA, a cancellation of matrix elements can occur for $\mu < 0$, allowing for the
cross sections to fall below the sensitivities of planned future detectors. This is exhibited
in Fig. 12, where $\sigma_{\tilde\chi^0_1-p}$ is plotted for $\tan\beta = 6$ (solid curve), 12 (dot-dash curve),
and 20 (dashed curve). (The $\tan\beta = 6$ curve terminates at low $m_{\tilde \chi^0_1}$ due
to the $m_h$ constraint, while the higher $\tan\beta$ curves terminate at low 
$m_{\tilde \chi^0_1}$ due to the
$b\rightarrow s \gamma$ constraint. The upper bound on $m_{\tilde \chi^0_1}$, corresponding to $m_{\tilde g}< $1
TeV, arises from the $\Omega h^2$ constraint.) One sees that the cross section goes through a
minimum at $\tan\beta \stackrel{\sim}{=}12$, though the expected rise at higher
 $m_{\tilde \chi^0_1}$ does not appear since the parameter space terminates before this sets
in.

\begin{figure}[htb]
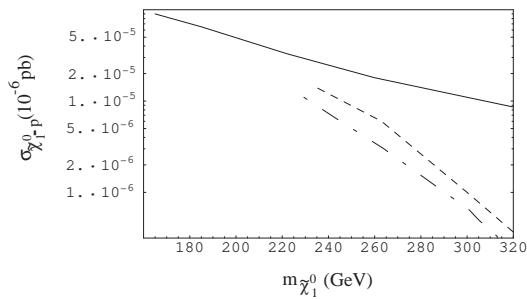

\centerline{ \DESepsf(adposmudark61215.epsf width 7 cm) }
\caption {\label{fig14} $\sigma_{\tilde\chi^0_1-p}$ for the D-brane model for $\mu < 0$ and
$\tan\beta = 6$ (solid), $\tan\beta = 12$ (dot-dash), and $\tan\beta = 15$ (dashed).}
\end{figure}

\section{Conclusions}

We have examined here the neutralino proton cross section for a number of
models possesing grand unification of the gauge coupling constants at 
$M_G\stackrel{\sim}{=}2\times10^{16}$ GeV. The mSUGRA model possesses universal supersymmetry soft
breaking masses. In the nonuniversal models, this universality is
maintained for the first two generations of squarks and sleptons and for
the gaugino masses, but relaxed for the third generation and the Higgs
scalar masses. In the D-brane models, universality is violated for both
squark and slepton masses at $M_G$ as well as for gaugino masses. Thus each
model has a different pattern of soft breaking. Our analysis here also
includes the recent experimental bounds on the Higgs mass of $m_h > 114$ GeV
\cite{lephiggs} (which for GUT models holds for all $\tan \beta$), and the
recent theoretical determination of the large $\tan\beta$
corrections to the NLO $b\rightarrow s\gamma$ decay \cite{gam,car}, both of which produce
significant effects on limiting the SUSY parameter space.
Despite the physical differences between the different models, the general
picture resulting is somewhat similar. Thus current detectors obeying Eq(1)
are sensitive to significant parts of the parameter space. For mSUGRA they
are sampling the regions where $\tan\beta \stackrel{>}{\sim} 25$. The nonuniversal models can
have cross sections a factor of 10 or larger (with an appropriate
choice of nonuniversalities) and so can sample the parameter space with
$\tan\beta \stackrel{>}{\sim} 7$. The D-brane models require $\tan\beta \stackrel{>}{\sim}$ 20.

Coannihilation effects play a crucial role for large $m_{\tilde{\chi}_1^0}$ in all the models,
and for large $\tan\beta$, they are sensitive to the value of $A_0$. Large $A_0$
leads to coannihilation corridors where $m_0$ can get quite large, thus
lowering the value of the $\sigma_{\tilde\chi^0_1-p}$ cross section. For $\mu > 0$, the cross sections
will generally still be accessible to planned future detectors obeying Eq.
(2), i.e.
\begin{equation}
              \sigma_{\tilde\chi^0_1-p}\stackrel{>}{\sim} 1\times10^{-10}\,
	      \,{\rm pb} \,\,\,{\rm for} \,\,\,m_{1/2} < 1\,{\rm TeV},\,\, \mu > 0
\end{equation}
However, in all models, a special cancellation of the Higgs amplitudes can
occur for $\mu < 0$  allowing the cross section to fall below the above bound
when $m_{1/2}\stackrel{>}{\sim} 450$ GeV. For mSUGRA, these cancellations produce minima where
the cross section essentially vanishes for a range of $m_{1/2}$ when 
$8 \stackrel{<}{\sim}
\tan\beta\stackrel{<}{\sim} 30$, for $m_{1/2} < 1$ TeV (see Fig.7) with similar results holding for
the nonuniversal models. The cancellations for the D-brane models occur for
$10\stackrel{<}{\sim}
\tan\beta\stackrel{<}{\sim}15$. We note that at fixed $\tan \beta$ the
cancellations can occur over a wide range of $m_{1/2}$, e.g. for mSUGRA $\tan \beta =
10$, $\sigma_{\tilde{\chi}_1^0-p}<10^{-10}$ pb for $400\, {\rm GeV} \leq m_{1/2}
\leq 1 \, {\rm TeV}$ \cite{note}. In such regions of parameter space, dark matter
detectors would not be able to observe Milky Way dark matter. However,
these regions of parameter space would imply that gluinos and squarks lie
above 1 TeV, but still should be accessible to the LHC if the parameter
space is bounded by $m_{1/2} < 1$ TeV. Thus other experimental consequences of
the models would be observable.


\begin{thebibliography}{99}
\bibitem{SUGRA}A.H. Chamseddine, R. Arnowitt and P. Nath,
\Journal{\PRL}{49}{970}{1982}; R. Barbieri, S. Ferrara and C.A. Savoy,
\Journal{\PLB}{119}{343}{1982}; L.
Hall, J. Lykken and S. Weinberg, \Journal{\PRD}{27}{2359}{1983}; P. Nath, R.
Arnowitt
and A.H. Chamseddine,  \Journal{\NPB}{227}{121}{1983}.

\bibitem{berezinsky}For previous analysis of nonuniversal models see:  V.
Berezinsky, A.
Bottino, J. Ellis, N. Fornengo, G. Mignola and S. Scopel, {\em Astropart. Phys.}
{\bf5}, 1 (1996);
{\em Astropart. Phys.} {\bf6}, 333 (1996); P. Nath and R. Arnowitt, 
\Journal{\PRD}{56}{2820}{1997};  R. Arnowitt and P. Nath,
\Journal{\PLB}{437}{344}{1998};  A. Bottino, F. Donato,
N. Fornengo and  S. Scopel, \Journal{\PRD}{59}{095004}{1999}; R. Arnowitt and P.
Nath, \Journal{\PRD}{60}{044002}{1999}.

\bibitem{munoz}L. Ibanez, C. Munoz and S. Rigolin,
\Journal{\NPB}{536}{29}{1998}.   

\bibitem{brhlik}M. Brhlik, L. Everett, G. Kane and J. Lykken,
\Journal{\PRD}{62}{035005}{2000}.

\bibitem{com}While the lower bound of Eq.(13) is somewhat lower than other
estimates, it allows us to consider the possibility that not all the dark matter
are neutralinos, i.e. the dark matter might be a mix of neutralinos, machos,
axions etc. Further, the minimum values of
$\sigma_{\tilde\chi^0_1-p}$ are not particularly sensitive to the lower bound
$\Omega_{\tilde\chi^0_1}h^2$.

\bibitem{fukugita}M. Fukugita,hep-ph/0012214.

\bibitem{gam}G. Degrassi, P. Gambino and G. Giudice, hep-ph/0009337 

\bibitem{car}M. Carena, D. Garcia, U. Nierste and C. Wagner, hep-ph/0010003 

\bibitem{bottino} A. Bottino et al. in ref.\cite{berezinsky}.

\bibitem{bottino1}A. Bottino, F. Donato, N. Fornengo and  S. Scopel,  {\em
Astropart. Phys.} {\bf 13}, 215 (2000).

\bibitem{falk}J. Ellis, T. Falk, K.A. Olive and  M. Srednicki,  {\em Astropart.
Phys.} {\bf 13}, 181 (2000).

\bibitem{ellis}J. Ellis, A. Ferstl and  K.A. Olive,
\Journal{\PLB}{481}{304}{2000}. 

\bibitem{falk1}J. Ellis, T. Falk, G. Ganis and K.A. Olive, 
\Journal{\PRD}{62}{075010}{2000}.

\bibitem{aads}E. Accomando, R. Arnowitt, B. Dutta and Y. Santoso,
\Journal{\NPB}{585}{124}{2000}.

\bibitem{aad}R. Arnowitt, B. Dutta and Y. Santoso,
 hep-ph/0005154.

\bibitem{nano}J. Ellis, G. Ganis, D. Nanopoulos and K. Olive, hep-ph/0009355. 

\bibitem{lephiggs}L3 Collaboration (M. Acciarri et al.). CERN-EP-2000-140,
 hep-ex/0011043; ALEPH Collaboration (R. Barate et al.). CERN-EP-2000-138,
  hep-ex/0011045. 

\bibitem{igo}P. Igo-Kemenes, talk presented at ICHEP 2000, Osaka, Japan, July
27-August 2, 2000.

\bibitem{lepchar}I. Trigger, OPAL Collaboration, talk presented at the DPF 2000, Columbus, OH;
T. Alderweireld, DELPHI Collaboration, talk presented at the DPF 2000, Columbus, OH.

\bibitem{cleo}M. Alam et al., \Journal{\PRL}{74}{2885}{1995}.

\bibitem{comm1} D0 Collaboration, \Journal{\PRL}{83}{4937}{1999}.

\bibitem{ellis2}J. Ellis and  R. Flores, \Journal{\PLB}{263}{259}{1991}; 
\Journal{\PLB}{300}{175}{1993}.

\bibitem{leutwyler}H. Leutwyler, \Journal{\PLB}{374}{163}{1996}.

\bibitem{mgo}M. Ollson, hep-ph/0001203. 

\bibitem{pas}M. Pavan, R. Arndt, I. Stravkovsky,
 and R. Workman, nucl-th/9912034, {\it Proc. of 8th International Symposium on
 Meson-Nucleon Physics and Structure of Nucleon}, Zuoz, Switzerland, Aug., (1999).

\bibitem{gs}J. Gasser and M. Sainio, hep-ph/0002283.

\bibitem{ads}R. Arnowitt, B. Dutta and Y. Santoso,
 hep-ph/0010244.
\bibitem{aadcp}E. Accomando, R. Arnowitt and  B. Dutta, \Journal{\PRD}{61}{075010}{2000}
\bibitem{neutmass}ALEPH Collaboration (R. Barate et al), hep-ex/0011047.
\bibitem{commx}The above LEP bound is model dependent and holds for the MSSM. We have checked,
however, that it still applies for the D-brane model.
\bibitem{note}We have considered here only the spin independent cross section.
As discussed in \cite{klapdor}, when the above cancelation is almost complete,
the true lower bound on $\sigma_{\tilde{\chi}_1^0-p}$ would be set by the spin
dependent part of the cross section. Precisely when this would occur depends on
the nuclei used in the target detector.

\bibitem{klapdor}V. Bednyakov and H. Klapdor-Kleingrothaus, hep-ph/0011233.

\end{thebibliography}
\end{document}